\documentclass[12pt,preprint]{aastex}

\usepackage{bm}
\usepackage{xcolor}
\usepackage{amsmath}
\usepackage{lineno}
\colorlet{myc1}{green!20!red!80!}
\definecolor{rkka}{RGB}{219,66,32}

\newcommand{\eb}{\begin{equation}}
\newcommand{\ee}{\end{equation}}

\shorttitle{Candidate exo-Jupiter in $\delta$ Pav}
\shortauthors{Makarov et al.}
\begin{document}

\title{Looking for astrometric signals below 20 m/s: A candidate exo-Jupiter in $\delta$ Pav} 
\author{Valeri V. Makarov, Norbert Zacharias, Charles T. Finch}
\affil{United States Naval Observatory, 3450 Massachusetts Ave. NW, Washington, DC 20392-5420, USA}
\email{valeri.makarov@gmail.com}

\begin{abstract}
We use a combination of Hipparcos space mission data with the USNO dedicated ground-based astrometric program URAT-Bright
designed to complement and verify Gaia results for the brightest stars in the south to estimate the small perturbations of
observed proper motions caused by exoplanets. One of the 1423 bright stars in the program, $\delta$ Pav, stands out
with
a small proper motion difference between our long-term estimate and Gaia EDR3 value, which corresponds to
a projected velocity of $(-17,+13)$ m s$^{-1}$. This difference is significant at a 0.994 confidence in the
RA component, owing to the proximity
of the star and the impressive precision of proper motions. The effect is confirmed by a comparison of long-term
EDR3-Hipparcos and short-term Gaia EDR3 proper motions at a smaller velocity, but with formally
absolute confidence. We surmise that the close Solar analog $\delta$ Pav harbors
a long-period exoplanet similar to Jupiter.

\end{abstract}


\section{Introduction}
\label{intr.sec}
As Jupiter orbits around the center of mass of the Solar system with a maximum orbital velocity of 13.7 km s$^{-1}$,
the Sun travels around the same point with an orbital velocity of 13 m s$^{-1}$. The Solar system barycenter is
located just outside the surface of the Sun, hence, the theoretically detectable reflex displacement is comparable to the apparent solar diameter.
For an observer at a distance of 10 pc from the Sun, the apparent semimajor axis of the Sun's reflex motion 
caused by the gravitational pull of Jupiter is not greater than
$0.5$ mas, which is hard to detect from Gaia observations alone. The formal uncertainty of Gaia
EDR3 proper motions, on the other hand, is already low enough to be looking for such signals from the nearest
stars. The detection method
based on the statistically significant difference between the long-term proper motions from absolute astrometric positions
separated by a sufficiently long time interval and precision short-term proper motions from a space mission,
called sometimes $\Delta\mu$-technique \citep{2005AJ....129.2420M, ker}, is presently the most sensitive with the 
available data. It allows us to
search for astrometric signals below 20 m s$^{-1}$, which are likely caused by distant giant planets similar to
Jupiter. 

In this Research Note, we are utilizing the results of a combined Hipparcos-URAT-UBAD astrometric solution
(HUU) obtained at USNO in order to supplement and verify Gaia data for the brightest stars
on the sky. The southern part of
this program called URAT-Bright was obtained with the USNO Robotic Astrometric Telescope \citep[URAT,][]{zac} from La Serena, Chile. The HUU catalog includes astrometric parameters of 1423 stars. The timeline of 22--26 years between the Hipparcos and URAT epochs provides precise proper motions
of comparable quality to those of Gaia EDR3 for these stars, which are very bright for Gaia. 
HUU proper motions are long-term and on the
Gaia coordinate system and 
independent of the short-term Gaia proper motions of bright stars.

\section{$\delta$ Pav, a nearby solar twin}
\label{star.sec}

$\delta$ Pav = HIP 99240 = HD 190248 is a bright southern star with a Gaia EDR3 parallax of $\varpi=163.95\pm 0.12$ mas,
which has been extensively observed in various photometric, spectroscopic, and astrometric programs. 
It is a close solar analog with a mass close to 1 solar mass as estimated from different models \citep{pin}
and an age of 5--6 Gyr. The estimated equatorial velocity and rotation period are also close to the solar values
\citep[e.g., $v\sin i=2.4$ km s$^{-1}$, $P_{\rm rot}=21.4$ d, ][]{hoj}. The metal content is slightly higher than the solar value
with different estimates ranging from [Fe/H]=0.31 to 0.37 \citep{net, cel}. The star does not have companions of
stellar mass, with observational limits in separation and magnitude difference ranging from
$\rho=0.03\arcsec$ at $\Delta$mag = 0 to $\rho=1.5\arcsec$ at $\Delta$mag = 3.0 
\citep{tok}. Based on the elemental abundances, the star is deemed to have a giant planet with a high probability
\citep{hin}.

\section{$\Delta\mu$ and its statistical significance}

Despite its brightness \citep[$V=3.536$ mag in the Geneva system,][]{net}, Gaia EDR3 provides accurate astrometric data
for this star. In particular, the proper motions for bright (problematic) stars have dramatically improved in
quality and precision with respect to Gaia DR2 \citep{lin}, making it possible to search for faint astrometric
signals from nearby exoplanets. The equatorial system proper motion from EDR3 at the mean epoch of 2016 is
$(1211.76\pm 0.07, -1130.24\pm0.10)$ mas yr$^{-1}$. The HUU proper motions are $(1211.16\pm 0.20, -1129.78\pm0.20)$ mas yr$^{-1}$.
The difference EDR3$-$HUU of $\Delta\mu=(0.601, -0.457)$ mas yr$^{-1}$ appears to be small, but it turns out to be
statistically significant with the formal errors and covariances. The simplest approach is to estimate the significance of
each $\Delta\mu$ component separately by computing the normalized squared difference with the formal variance in the
denominator being the sum of the corresponding variances in HUU and EDR3. The resulting squared normalized component differences are $(7.994,4.127)$. They are supposed to be distributed
as $\chi^2(1)$ and the cumulative distribution function (CDF) values are CDF$_{\chi^2}(7.994,4.127)=(0.995,0.958)$. 
Thus, there is
a negligibly small probability that the derived $\Delta\mu$ can be explained by random observational error, as long as the
estimated uncertanties are valid. The corresponding perturbation of tangential velocity of the star is
 $(17.4,-13.2)$ m s$^{-1}$, which are consistent with the values expected from a Jupiter
analog orbiting $\delta$ Pav.

The URAT part of observational data can be compromised by unaccounted instrumental effects, although the star is quite
``clean" from the astrometric point of view and there are no bright neighbors on the sky. To verify the presence of a
small perturbation in its apparent trajectory, we re-derived the $\Delta\mu$ vector using a different approach.
The long-term proper motion is computed directly from the difference in the mean positions in Hipparcos (mean epoch
1991.25) and Gaia EDR3 (epoch 2016). It is assumed in the computation of associated covariances that the Hipparcos position and EDR3
proper motion are statistically independent.

\begin{figure}[htbp]
  \centering
  \plotone{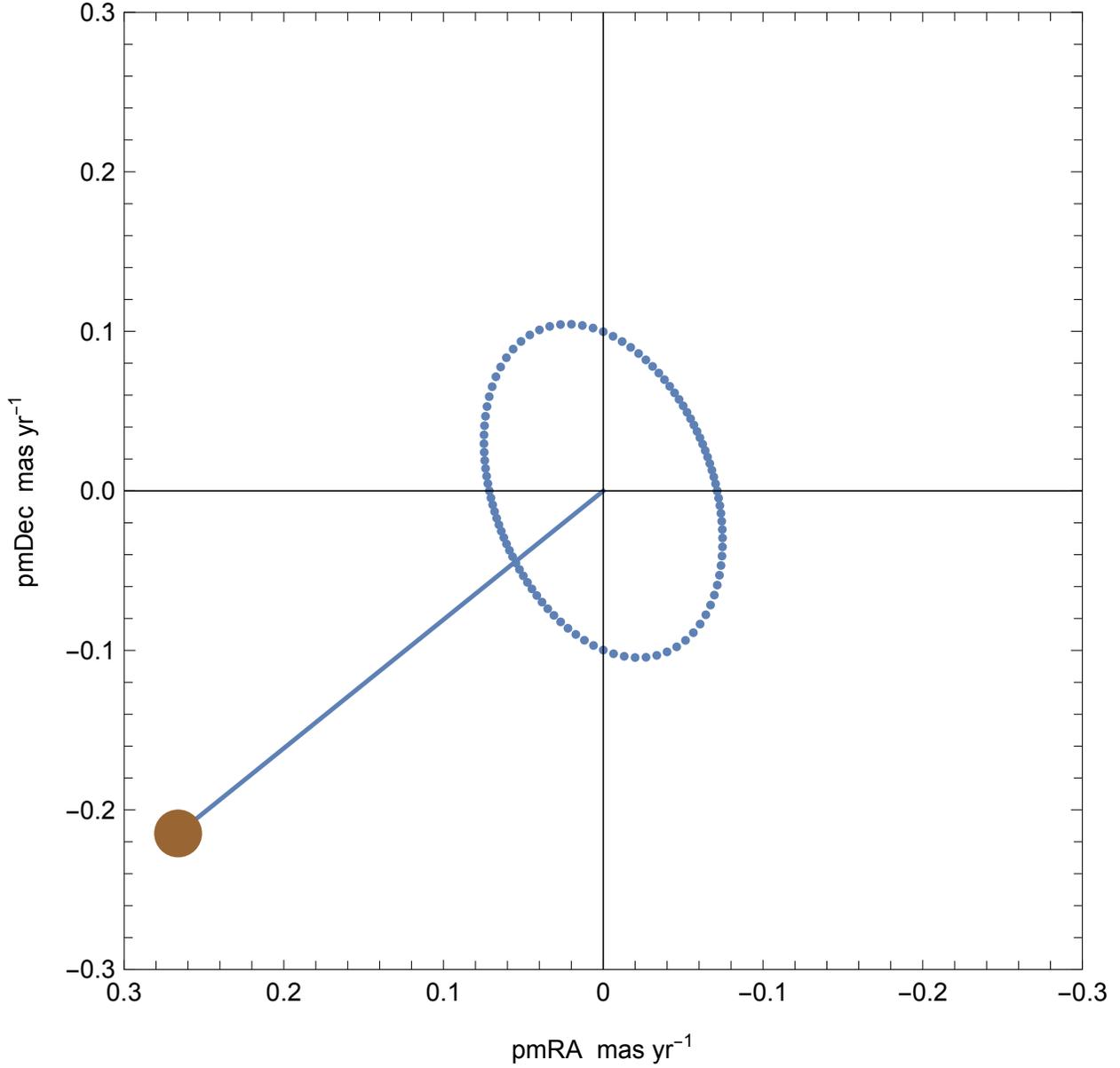}
\caption{Proper motion difference vector EDR3$-$HG for $\delta$ Pav marked with the large brown circle at the end point
and the corresponding error ellipse of this vector in the equatorial sky tangent plane. The vector is roughly aligned with the 
minor axis of the error ellipse, which creates most favorable conditions for detecting small signals. \label{ellipse.fig}}
\end{figure}

Fig. \ref{ellipse.fig} shows the error ellipse and the actual $\Delta\mu$ vector computed from Hipparcos and Gaia data only.
The proper motion difference is $\Delta\mu=(0.266, -0.215)$ mas yr$^{-1}$. For the computation of the error ellipse,
cf., for example, \citep{mak17}. Using the covariance matrix, we compute the formal variance of $\Delta\mu$ norm
in the given direction of the vector. The ratio of the squared norm and its formal variance is expected to be distributed as
$\chi^2(2)$, and the CDF of this distribution at the estimated value yields the confidence level of detection, which is
the probability that the observed value is not caused by purely random error. We obtain a confidence level of 0.999.
The corresponding tangential velocity signal is $(7.7,-6.2)$ m s$^{-1}$.

\section{Conclusions}
\label{con.sec}
Using two somewhat different techniques and data sets, we derived the proper motion difference vectors between
long-term data, where the reflex orbital motion caused by a Jupiter-like exoplanet is likely to be much
reduced, and the short-term Gaia EDR3 proper motions, where this effect is averaged out to a lesser degree.
Both techniques indicate the presence of a small astrometric signal at very high (formal) confidence levels.
The values of $\Delta\mu$, however, are numerically different with the less precise URAT estimation
showing a larger effect. The signs of proper motion components are the same, as well as the general direction.
We speculate that this signal may be caused by a widely separated, long-period giant planet similar to our Jupiter.

\end{document}